\documentclass[11pt]{article}
\usepackage[dvipsnames,x11names]{xcolor} 
\usepackage{lscape}
\usepackage{amssymb} 
\usepackage{graphicx}
\usepackage{footnote}
\usepackage[utf8]{inputenc}	
\usepackage{mathtools}
\usepackage{amsthm}
\usepackage{amsfonts}
\usepackage[english]{babel}
\usepackage{bigints}
\usepackage{mathrsfs}
\usepackage{lineno}
\usepackage{nicefrac}
\usepackage{enumerate}
\usepackage{bbm}

\usepackage{listings}

\usepackage{commath}
\usepackage{multirow}

\usepackage{newtxtext}
\usepackage[T1]{fontenc}

\usepackage{algorithm} 
\usepackage{algpseudocode} 
\usepackage{booktabs}

\usepackage[most]{tcolorbox}
\usepackage{mdframed}

\usepackage[]{natbib}
\usepackage{csquotes}

\usepackage{tabularx}

\usepackage[colorlinks=true,citecolor=blue,linkcolor=black,urlcolor=blue]{hyperref}
\definecolor{light-gray}{gray}{0.91}

\usepackage[margin=2cm]{geometry}

\usepackage{lineno}

\usepackage{sectsty}
\sectionfont{\color{BrickRed}}

\sectionfont{\color{RoyalBlue4}}
\subsectionfont{\color{RoyalBlue4}}
\subsubsectionfont{\color{RoyalBlue4}}
\paragraphfont{\color{RoyalBlue4}}

\allowdisplaybreaks[1]
\usepackage{pifont}

\makeindex

\def\tr{\textrm{tr }}

\def\dvg{\textrm{Div}}

\def\crl{\textrm{Curl }}

\def\sk{\text{skew }}
\def\sy{\text{sym }}

\def\dev{\textrm{dev }}

\newcommand{\Lagr}{\mathcal{L}}

\newcommand{\Cmacro}{\mathbb{C}_{\rm macro}}
\newcommand{\Crelax}{\mathbb{C}_{\rm relax}}
\newcommand{\Crel}{\mathbb{C}_{\rm rel}}
\newcommand{\Crot}{\mathbb{C}_{\rm rot}}
\newcommand{\Cspin}{\mathbb{C}_{\rm spin}}

\newcommand{\mumacro}{\mu_{\rm macro}}
\newcommand{\murel}{\mu_{\rm rel}}
\newcommand{\murot}{\mu_{\rm rot}}
\newcommand{\muspin}{\mu_{\rm spin}}

\newcommand{\lambdamacro}{\lambda_{\rm macro}}
\newcommand{\lambdarel}{\lambda_{\rm rel}}
\newcommand{\lambdarelax}{\lambda_{\rm relax}}

\newcommand{\murelax}{\mu_{\rm relax}}

\newcommand{\CCmacro}{\mathsf{C}_{\rm macro}}
\newcommand{\CCe}{\mathsf{C}_{\rm e}}
\newcommand{\CCmicro}{\mathsf{C}_{\rm relax}}

\newcommand{\mumicro}{\mu_{\rm relax}}

\newcommand{\lambdamicro}{\lambda_{\rm relax}}
\newcommand{\Ce}{\mathbb{C}_{\rm e}}

\newcommand{\mue}{\mu_{\rm e}}

\newcommand{\lambdae}{\lambda_{\rm e}}

\newcommand{\Czero}{\mathbb{C}_0}
\newcommand{\Cinfty}{\mathbb{C}_{\infty}}

\newcommand{\id}{{\boldsymbol{\mathbbm{1}}}}

\begin{document}


\begin{flushleft}

	{\huge\bfseries \color{DodgerBlue4}{A micromorphic constitutive framework for (non)linear softening modeling}\par}

	\vspace{1.5em}  
	
	Rafael Abreu\textsuperscript{1,*}\\[0.5em]  
	
	\scriptsize 
	\textsuperscript{1} \textit{Institut de Physique du Globe de Paris, CNRS, Universit\'e de Paris, Paris, France} \\ [1em]  
	\textsuperscript{*} {corresponding author: rabreu@ipgp.fr}
	
	\normalsize 
\end{flushleft}

%
%

\begin{abstract}

	Nonlinear shear softening in soils and rocks is commonly described using empirical modulus-reduction laws. In this work, we investigate nonlinear softening within the Eringen--Mindlin micromorphic framework by interpreting its additional constitutive parameters in terms of coupling, internal relaxation, and micro-inertia. The micromorphic continuum is viewed as a dynamically enriched effective medium in which the macroscopic deformation interacts with an internal microstructural deformation field. This interaction naturally produces frequency-dependent effective elastic properties and provides a physical connection between macroscopic wave propagation and unresolved microstructural dynamics.
	
	We show that nonlinear shear softening can be described by a progressive reduction of the coupling between macroscopic and internal deformation. As this coupling weakens with increasing shear strain, the effective shear modulus decreases. In the low-frequency limit, the resulting constitutive relation takes a hyperbolic form closely related to the classical Hardin--Drnevich law. The model is validated against laboratory measurements of shear-modulus reduction in sand and reproduces the observed nonlinear behavior with only a small number of physically interpretable parameters.
	
	The proposed framework provides a new description of nonlinear softening and frequency-dependent behavior without requiring explicit knowledge of the underlying microstructural geometry. Classical empirical modulus-reduction curves emerge as a macroscopic consequence of the interaction between macroscopic and internal deformation.
	
\end{abstract}

%


\newpage
\normalsize
\section{Introduction}

Micro-continuum theories \citep{mindlin1964,Eringen1999} provide a natural framework to extend classical linear elasticity by incorporating additional internal degrees of freedom. The micro-continuum field theories cover a wide range of three main different theories: micropolar, microstretch and micromorphic media \citep{Eringen1999}. The theory of Cosserat/micropolar media \citep{Cosserat1909}, describes a continuum material with additional free rotational degrees of freedom at each location \citep{Eringen1976,Nowacki1986,jeong2008existence,neff2009new,Kulesh2009,grekova2009waves,Grekova2012A,Grekova2012B,Grekova2016}. Each particle inside the material can independently rotate. The theory of microstretch media describes elastic materials with rotational and stretching free degrees of freedom \citep{Eringen1999}. Each particle inside the material can only independently rotate and have stretching (breathing) motions. The theory of micromorphic media describes a continuum that allows any kind of affine micro-distortion. While these models promise a powerful description of media with microstructure, their practical application is often limited by the large number of constitutive parameters and the difficulty of assigning them a clear physical meaning.

The application of micro continuum field theories in the Earth sciences is not new \citep[e.g.][]{Teisseyre1973,teisseyre2006earthquake,Nagahama2000,Nagahama2001425,Teisseyre2008_2,Teisseyre2008,Teisseyre01082011}. Pioneering work by \cite{Teisseyre1973,Teisseyre1974} introduced a simplified general micromorphic model to describe earthquake processes, based on a symmetric microstrain formulation, as a particular case of the general micromorphic theory proposed by \cite{eringen1970micromorphic}. However, due to its symmetric formulation, this early model is unable to represent micro-rotational deformations. This limitation motivates the use of more general formulations, such as the relaxed micromorphic model, in which both symmetric and antisymmetric microstructural effects are effectively modeled.

A different model has been introduced by \cite{teisseyre2009fundamental} in order to describe the earthquake source processes. This model relies on the consideration of a different perspective of motions and deformations in an asymmetric continuum, including simple motions (translation and rotation, named spin) and simple deformations. It is based on the constitutive relation introduced by  \cite{Shimbo1975} joining the rotations with the antisymmetric stresses \citep{Teisseyre2007_Abstract}\footnote{the independent rotation field and related constitutive law joining the antisymmetric stresses and rotations, e.g., rotations related to grains (the points of continuum), have been considered by \cite{Shimbo1975} in his considerations on the friction and fracturing processes. In this way, there arose an idea to construct the theory of a continuum with asymmetric force stresses.}.

Despite these advances in the Earth sciences, the role of microstructure in seismic wave propagation remains still poorly understood. In particular, classical approaches to account for microstructural effects in seismology are typically based on homogenization theory, in which small-scale heterogeneities are averaged out to obtain effective macroscopic properties \citep[e.g.][]{capdeville20102,capdeville2013residual,capdeville2015fast,capdeville2018elastic,hedjazian2021multiscale}. While these methods provide a rigorous link between microscale structure and large-scale behavior under conditions of strong scale separation, they lead to effective models in which the microstructure no longer appears explicitly. As a consequence, the resulting description is purely macroscopic and does not retain internal degrees of freedom that could participate dynamically in wave propagation. As we will see, this limitation becomes particularly relevant when the microstructure influences wave propagation behavior within an observable frequency range. In these situations, seismic observables such as velocity dispersion, modulus reduction, and frequency-dependent responses cannot be directly linked to underlying physical mechanisms, since the microstructure has been entirely eliminated from the mathematical description. As a result, different physical processes may lead to indistinguishable macroscopic behavior, limiting the interpretability of seismic observations. In such cases, an alternative modeling strategy is required, in which microstructural dynamics are not eliminated but incorporated in the mathematical description of the wave dynamics \citep{Eringen1999}.

We investigate linear elastic wave propagation within the Eringen-Mindlin micromorphic framework, with a particular focus on a new physical interpretation of its constitutive parameters and realistic Earth sciences applications. We show that the required micromorphic model parameters characterize the fundamental physics that governs micro and macro scales, providing a direct fundamental physical description of the underlying dynamical mechanisms. Within this perspective, macroscopic wave behavior naturally emerges from the interaction between elastic response and internal relaxation mechanisms.

This paper is now organized as follows. In Section \ref{sec.Equations_of_Motion} we present the micromorphic theoretical framework within a clear notation and with a particular emphasis to the physical interpretation of the model parameters. In Section \ref{se_GeneralExperimentalValidations}, we validate the micromorphic model against experimental and geophysical observations in experiments of strain-dependent shear softening. Section~\ref{sec.Discussion} provides a broader interpretation of the results. We further discuss connections with homogenization theory. Finally, conclusions are drawn in Section~\ref{sec.Conclusions}.

\section{Theoretical foundations}
\label{sec.Equations_of_Motion}

The equation of motion of any continua can be found, in a simple way, by using the Lagrangian formalism also called the variational principle of least action and/or the principle of stationary action. The Lagrangian formalism aims to find the extremal functional(s) that make the functional $S$ obtain a minimum value (principle of stationary action $\delta S = 0)$. The functional $S$ can be written in general as follows
\begin{align}
	S = \int_0^t \int _{\Omega} \Lagr \dif x \dif t = \int_0^t \int _{\Omega} (K - \mathcal{E}) \dif x \dif t ,
	\label{eq.Minimization_Functional}
\end{align}
were $\Omega$ denotes the volume of the continuum,  $\Lagr$ is the Lagrangian defined as the difference between kinetic energy $K$ and potential energy $\mathcal{E}$. The functional(s) that minimizes $S$ is simply found by setting $\delta S = 0$. 

It is thus straightforward to find equations of motion in continuum media including micro-structure (or not), by defining the appropriate kinetic $K$ and potential (internal elastic free) $\mathcal{E}$ energies. The potential energy $\mathcal{E}$ is of main importance, since depending on how we define $\mathcal{E}$, the definition of the kinetic energy is straightforward in a first attempt.

\subsection{Internal free energy}

In conventional linear elastic media, the potential energy $\mathcal{E}$ is defined as follows (see \cite{Slawinski2015} for further details)
\begin{align}
	\mathcal{E} (\varepsilon) =  \frac{1}{2} \varepsilon_{ij} \, \mathbb{C}_{ijkl} \, \varepsilon_{kl}, \qquad \text{with} \qquad  \varepsilon_{ij} = \partial_j u_i + \partial_i u_j ,
	\label{eq.General_Linear_Elastic_Internal_Energy}
\end{align}
where $u:\mathbb{R}^3 \rightarrow \mathbb{R}^3$ denotes the macro-displacement and $\mathbb{C}_{ijkl}$ is a fourth-order tensor of elastic constants. In the rest of this paper we will denote the fourth-order tensor of elastic constants related to the linear elasticity as  
\begin{align}
	(\Cmacro)_{ijkl} = \mathbb{C}_{ijkl} .
	\label{eq.Cmacro_definition}
\end{align}
Note that the strain tensor $\varepsilon_{ij}$ is symmetric $(\varepsilon_{ij}=\varepsilon_{ji})$ and it does not involve any other physical quantities than the displacement $u$. In contrast, the general internal elastic free energy for the Eringen-Mindlin micromorphic continuum can be written as follows 
\begin{align}
	\begin{aligned}
		\mathcal{E} (\varepsilon,\epsilon,e,\gamma) =  \underbrace{\frac{1}{2}\epsilon_{ij} \, \mathbb{A}_{ijkl} \, \epsilon_{kl}}_{\text{elastic relative free energy}}  + \underbrace{\frac{1}{2} e_{ij} \, \mathbb{B}_{ijkl} \, e_{kl}}_{\text{relaxation free energy}}  +  \underbrace{\frac{1}{2} \gamma_{klm} \,\mathbb{L}_{klmnpq} \, \gamma_{npq}}_{\text{ curvature energy}} ,
	\end{aligned}
	\label{eq.General_Eringen_Micromorphic_Internal_Energy}
\end{align}
where $\mathbb{A},\mathbb{B}$, are fourth-order tensors and $\mathbb{L}$ a six-order tensor of curvature elastic constants \citep{Eringen1999}. The strain measures $ (\varepsilon,\epsilon,e,\gamma)$ are defined by the following expressions
\begin{align}
\begin{aligned}
	 \epsilon_{kl} & =  \partial_k u_l  - P_{kl}, \qquad  e_{kl}  =  P_{kl},  \qquad  \gamma_{klm} = \partial_m P_{kl}  .
\end{aligned}
\end{align}
In a symbolic formulation we write
\begin{align}
	\begin{aligned}
		\mathcal{E} (\varepsilon,\epsilon,e,\gamma) & = \left \langle (\nabla u-P) , \mathbb{A} (\nabla u-P)\right \rangle   + \left \langle P , \mathbb{B} P\right \rangle +  \left \langle \nabla P , \mathbb{L} \nabla P\right \rangle, \\
		& = \left \langle \sy (\nabla u-P) , \Crel \, \sy (\nabla u-P)\right \rangle   + \left \langle \sk (\nabla u-P) , \Crot \,\sk (\nabla u-P)\right \rangle \\
		& + \left \langle \sy P , \Crelax \, \sy P\right \rangle +  \left \langle \sk P , \Cspin \, \sk P\right \rangle + \left \langle \nabla P , \mathbb{L} \nabla P\right \rangle,
		\label{eq.General_Eringen_Energy}
	\end{aligned}
\end{align}
where $u:\mathbb{R}^3 \rightarrow \mathbb{R}^3$ denotes the usual macro-displacement and $P:\mathbb{R}^3 \rightarrow \mathbb{R}^{3\times 3}$ denotes the micro-distortion second-order tensor allowing for the extra degrees of freedom: the skew-symmetric part---$\sk P$---represents independent rotational degrees of freedom, the spherical part---$\frac{1}{3} \tr (P)$---represents possible breathing motions and the deviatoric symmetric part---$\dev \sy P$---involves independent stretch (volume preserving) motions. The terms $\Crel,\Crot,\Crelax,\Cspin$ are fourth-order tensors of elastic constants.

One can note that the internal elastic energy for the Eringen-Mindlin micromorphic model (eq. \eqref{eq.General_Eringen_Micromorphic_Internal_Energy}) is much more complex compared to the linear elastic case (eq. \eqref{eq.General_Linear_Elastic_Internal_Energy}). The main and most important difference is that the strain at the macro-structural level given by---$\nabla u$---is different from the strain at the local-structural level---$P$. 

Formally, the resulting stress measures are then given by the following expressions
\begin{align}
	\begin{aligned}
		\sigma_{ij}& = \partial_{\epsilon} \mathcal{E} = \mathbb{A}_{ijkl} \, \epsilon_{kl} = (\Crel)_{ijkl} \, \sy \epsilon_{kl} + (\Crot)_{ijkl} \, \sk \epsilon_{kl} , \\
		s_{ij} & = \partial_{e} \mathcal{E}  = \mathbb{B}_{ijkl} \, e_{kl} = (\Crelax)_{ijkl} \, \sy e_{kl} +  (\Cspin)_{ijkl} \, \sk e_{kl}  ,\\
		m_{klm} & = \partial_{\gamma} \mathcal{E} = \mathbb{L}_{klmnpq}  \, \gamma_{npq} , 
		\label{eq.Stress_Measures_Eringen_Micromorphic}
	\end{aligned}
\end{align}
where $\sigma$ is called the force-stress tensor, $m$ is called stress-moment tensor and $s$, we call the relaxation-stress tensor, and the supper indices \textit{sym} and \textit{skew} refer to the symmetric  and anti-symmetric parts respectively. We can explicitly write the general resulting stress measures as follows
\begin{align}
	\begin{aligned}
		\sigma & =  \Crel  \, \sy (\nabla u - P)  + \Crot \, \sk (\nabla u - P) , \\
		s & =  \Crelax \, \sy P + \Cspin \, \sk P, \\
		m & =  L_c^2   \, \Cmacro \, \nabla P,
		\label{eq.Stress_Measures_Eringen_Micromorphic2}
	\end{aligned}
\end{align} 
where, for simplicity, we have considered $\mathbb{L}= L_c^2 \, \Cmacro$. The isotropic stress measures corresponding to eqs. \eqref{eq.Stress_Measures_Eringen_Micromorphic2} are given by the following expressions
\begin{align}
	\begin{aligned}
		\sigma & = 2 \murel \, \sy (\nabla u - P) + 2\murot \, \sk (\nabla u - P)  + \lambdarel \, \tr (\nabla u - P) \cdot \id ,\\
		s & = 2 \murelax \, \sy P + 2 \muspin \, \sk P  + \lambdarelax  \tr (P) \, \id ,\\ 
		m & =  L_c^2  \,\mumacro  \, \nabla P ,
	\end{aligned}
\end{align}
where the parameter $\mumacro$ is having the role of the classical Lam\'e shear modulus. A new set of elastic constants $\lambdarel,\lambdarelax,\murel,\murelax,\muspin\geq 0$ are introduced and need to be determined. The elastic constant $\murot$ is called the \textit{Cosserat couple modulus}. Finally $L_c$ is a characteristic length  related to the problem in study.

\subsection{Kinetic energy and equations of motion}

Having defined the elastic free energy eq. \eqref{eq.General_Eringen_Micromorphic_Internal_Energy}, it is straightforward to formally define the corresponding kinetic energy as follows
\begin{align}
	K = \frac{1}{2} \rho \, \partial_t u \, \partial_t u + \frac{1}{2} \eta \, \partial_t P \, \partial_t P,
	\label{eq.Relaxed_Micromorphic_Kinetic_Energy}
\end{align}
where $\rho$ is the macroscopic scalar density, $\eta$ is the micro-inertia density tensor of the micro-distortion (in the isotropic case simply a scalar). Solving the stationary condition problem $\delta S = 0$, with $S$ defined by eq. \eqref{eq.Minimization_Functional}, we can find the general, second-order coupled, micromorphic equations of motion \citep{Eringen1999} 
\begin{align}
\begin{aligned}
		\rho \,  \partial_t ^2 u  & = \dvg\left[\Crel  \, \sy (\nabla u - P) + \Crot \, \sk (\nabla u - P)  \right],  \\
		 \eta \, \partial_t ^2P  &= \dvg \left[L_c^2  \,\Cmacro  \, \nabla P\right] + \left[\Crel  \,\sy  (\nabla u - P)  + \Crot \, \sk (\nabla u - P)\right] \\
		 & - \Crelax \, \sy P - \Cspin \sk P, 
\end{aligned}
	\label{eq.Relaxed_Micromorphic_Final_Eq_Motion_No_Source}
\end{align}
with the free surface boundary conditions given by
\begin{align}
	\sigma \cdot\hat{n} = 0 \quad \mbox{on} \quad  \partial \, \Omega, \qquad  m \cdot\hat{n} = 0 \quad \mbox{on} \quad \partial \, \Omega ,
\end{align}
where $\hat{n}$ refers to the direction normal to the surface $\partial \, \Omega$.

\subsection{The relaxed micromorphic model}

The relaxed micromorphic model introduced in \cite{Neff2014} attempts to reduce the number of parameters required by the original Eringen/Mindlin model and to provide clarity in the application of micromorphic theory. Within the relaxed micromorphic framework, the number of independent parameters is significantly reduced, while retaining the ability to describe nonlocal effects, dispersion, and scale-dependent behavior. It has been recently applied in the engineering field for the design of seismic metamaterials \citep[e.g.][]{Madeo_bandgaps_2015,Madeo2016a,Madeo2016b,hermann2024design,demetriou2025effective,madeo2017modeling,hermann2026unveiling,barbagallo2019relaxed,rizzi2024frequency,d2020effective,voss2023remarks,ramirez2024effective,Madeo2016a,Madeo2016b,neff2020identification}.

The general elastic energy is given by the following expression
\begin{align}
	\begin{aligned}
		\mathcal{E} (\varepsilon,\epsilon,e,\gamma) & = \left \langle (\nabla u-P) , \mathbb{A} (\nabla u-P)\right \rangle   + \left \langle \sy P , \mathbb{B} \, \sy P\right \rangle + \left \langle \crl P , \mathbb{L} \, \crl P\right \rangle, \\
		& = \left \langle \sy (\nabla u-P) , \Crel \, \sy (\nabla u-P)\right \rangle   + \left \langle \sk (\nabla u-P) , \Crot \,\sk (\nabla u-P)\right \rangle \\
		& + \left \langle \sy P , \Crelax \, \sy P\right \rangle + \left \langle \crl P , \mathbb{L} \, \crl P\right \rangle,
		\label{eq.General_Rlaxed_Energy}
	\end{aligned}
\end{align}

The main conceptual difference with the original Eringen--Mindlin model lies in relaxing the requirement that the full gradient of the micro-deformation, $\nabla P$, be energetically penalized. The underlying idea is that not all components of $\nabla P$ are required to capture the mechanically relevant nonlocal effects. Therefore, the curvature energy is formulated in terms of reduced differential operators acting on $P$, such as $\crl P$, $\dvg P$, or any other combination. Thus the relaxed micromorphic model corresponds to any of those particular choices \citep[e.g.][]{Madeo2016b}.

However, it is important to emphasize that the decomposition of the relative deformation into its symmetric and skew-symmetric parts,
\begin{align}
	\nabla u-P
	=
	\sy(\nabla u-P)
	+
	\sk(\nabla u-P),
\end{align}
is not unique to the relaxed micromorphic model. Rather, the same decomposition follows naturally from the general Eringen--Mindlin formulation and is therefore equally valid within the original Eringen--Mindlin framework, independently of the particular choice of curvature energy.

\subsection{Main contributions of this work}

For the purposes of the present work, we adopt the constitutive framework of the Eringen--Mindlin micromorphic theory while expressing the relative deformation energy through its decomposition into symmetric and skew-symmetric parts. When the curvature energy is neglected, i.e., $\nabla P=\crl P=0$, the remaining local elastic energy is identical in both the Eringen--Mindlin and relaxed micromorphic formulations.

The objective of the present work is not to propose a new micromorphic constitutive model, but rather to provide a new physical interpretation of the classical Eringen--Mindlin formulations and to validate it against experimental observations. 

While the governing equations of Eringen--Mindlin formulation remain unchanged, the constitutive tensors are reinterpreted from a new mechanical perspective. To the best of the author's knowledge, this physical interpretation has not been presented previously.

The remainder of this work we adopt the notation $\Crel$ and $\Crot$ to denote the energetic contributions associated with the symmetric and skew-symmetric parts of the relative deformation tensor $(\nabla u-P)$. This notation is used exclusively to facilitate the physical interpretation developed in the following sections.

In particular, the tensors $\Crel$, $\Crot$, $\Crelax$, and $\Cspin$ are interpreted as energetic resistances associated with distinct deformation mechanisms rather than merely as constitutive coefficients. As we will see, within the introduced framework, $\Crel$ and $\Crot$ govern the resistance to the relative deformation between the macroscopic and microstructural fields, whereas $\Crelax$ and $\Cspin$ quantify the energetic resistance to the activation of internal deformation modes. 

This new interpretation naturally leads to a dynamic homogenization viewpoint, in which the effective constitutive response emerges from the interaction between the macroscopic deformation and an local/microstructural deformation. As a consequence, the constitutive parameters acquire a direct mechanical interpretation that, to the best of the author's knowledge, is absent from the original Eringen--Mindlin and relaxed micromorphic formulations.

\subsection{Normal mode solutions}
\label{app.Consistent_BC}

Considering $L_c=0$, we can write the general equations of motion (eq. \eqref{eq.Relaxed_Micromorphic_Final_Eq_Motion_No_Source}) as follows
\begin{align}
	\begin{aligned}
		\rho \,  \partial_t ^2 u  & = \dvg\left[\Crel  \, \sy (\nabla u - P)+ \Crot \, \sk (\nabla u - P) \right],  \\
		\eta \,  \partial_t ^2 P  &= \left[\Crel  \, \sy (\nabla u - P) + \Crot \, \sk (\nabla u - P) \right] - \Crelax \, \sy P  - \Cspin \, \sk P . 
	\end{aligned}
	\label{eq.General_Microbalance_Equation}
\end{align}
We next assume normal modes solutions of the form
\begin{align}
	u(x,t) = \varphi(x) \, e^{i\omega t}, \qquad P(x,t) = \Psi(x) \, e^{i\omega t},
	\label{eq.normal_modes_solutions}
\end{align}
where $\varphi:\Omega \to \mathbb{R}^3$ and $\Psi:\Omega \to \mathbb{R}^{3\times 3}$ are spatial eigenfunctions defined on the domain $\Omega$, and $\omega$ denotes the associated eigenfrequency. 

Substitution of the normal mode solutions (eq. \eqref{eq.normal_modes_solutions}) into the governing equations of motion (eq. \eqref{eq.Relaxed_Micromorphic_Final_Eq_Motion_No_Source}), leads to a coupled eigenvalue problem for $(\varphi,\Psi)$ and $\omega^2$. From the second equation in eq. \eqref{eq.General_Microbalance_Equation}, the independent deformation field $\Psi$ is linearly related to the macroscopic deformation gradient $\nabla \varphi$ as follows
\begin{align}
	\begin{aligned}
		\left[\left(\Crel+\Crelax\right)- \eta \, \omega^2 \id \right]  \sy \Psi  & =   \Crel \, \sy (\nabla \varphi) ,  \\
		\left[\left(\Crot+\Cspin \right)- \eta \, \omega^2 \id  \right] \sk \Psi  & = \Crot \, \sk (\nabla \varphi) ,
		\label{eq.General_Normal_modes_equations_P_nablau}
	\end{aligned}
\end{align}
where $\id$ denotes the fourth-order identity tensor. If we insert eq. \eqref{eq.General_Normal_modes_equations_P_nablau} into the first equation of eq. \eqref{eq.General_Microbalance_Equation}, and consider the anti-symmetric part to be zero, we obtain the effective condition
\begin{align}
	 \Cmacro 
	& =
	\Crel
	-
	\Crel 
	\left[ \Ce + \Crelax - \eta \, \omega^2 \id \right]^{-1}
	\Crel .
	\label{eq.Cmacro_effective}
\end{align}

The effective elasticity tensor $\Cmacro$ (defined in eq. \eqref{eq.Cmacro_definition}) is frequency dependent and it possesses a pole whenever
\begin{align}
	\Crel+\Crelax-\eta \, \omega^2\id=0,
\end{align}
for which the expression is no longer well defined. At these frequencies the constitutive model is no longer well defined, and dissipative effects must be incorporated.

\begin{figure}
	\begin{center}
		\includegraphics[width=0.8\textwidth]{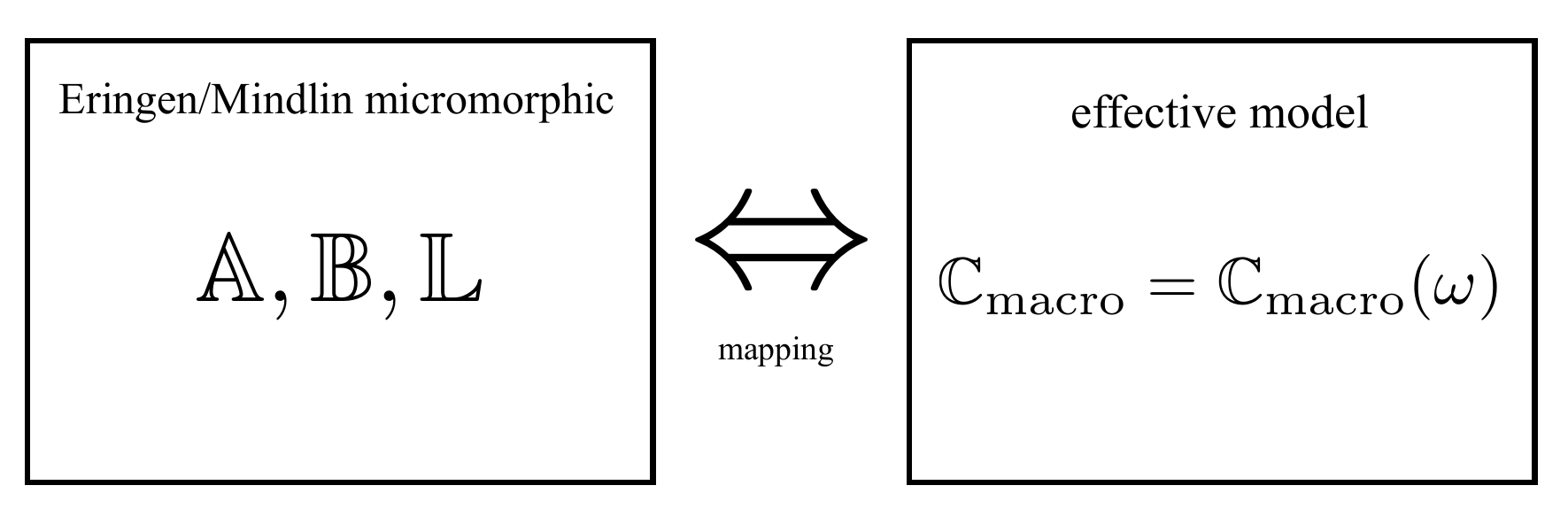}
		\caption{Micromorphic mapping to an linear elastic model with frequency dependent elastic parameters.}
		\label{Fig.effective_mapping}
	\end{center}
\end{figure}

Equation \eqref{eq.Cmacro_effective} simply shows that we are able to map the micromorphic model to an effective linear elastic model with frequency dependent parameters (see Fig. \ref{Fig.effective_mapping}).
We can observe that the effective frequency dependent macroscopic response $(\Cmacro=\Cmacro(\omega))$ emerges from the interplay between the elastic stiffness $\Crel$, the relaxation stiffness $\Crelax$, and the dynamic contributions associated with micro-inertia term $\eta \,  \omega^2$. As a result, the effective elastic response becomes frequency dependent, that can be understood as emergent form the general microstructural internal dynamics.

In the limiting case where the micro-inertia is negligible, i.e., $\eta \to 0$, the dynamic contributions vanish and the static consistency condition is recovered 
\begin{align}
	\Cmacro = \Crel \, \Crelax \left(\Crel + \Crelax\right)^{-1}.
	\label{eq.Consistency_Condition}
\end{align}

Using the skew-symmetric part of the equation of motion (eq. \eqref{eq.General_Microbalance_Equation}), and eq. \eqref{eq.General_Normal_modes_equations_P_nablau} we can write
\begin{align}
	\Crot \, \sk(\nabla\varphi-\Psi)
	=
	\left(
	\Cspin-\eta\, \omega^2\id
	\right)\sk\Psi .
\end{align}
Since the effective macroscopic elasticity tensor $\Cmacro$ is symmetric, the skew-symmetric contribution must vanish. This is satisfied either when $\sk\Psi=0$ or when
\begin{align}
	\Cspin-\eta \, \omega^2\id=0.
\end{align}

This condition represents a dynamic balance between the rotational elastic stiffness $\Cspin$ and the micro-inertia $\eta$. Under this condition, the skew-symmetric part of the relative deformation no longer contributes to the macroscopic stress, i.e.,
\begin{align}
	\Crot\,\sk(\nabla\varphi-\Psi)=0,
\end{align}
so that the effective macroscopic constitutive response is purely symmetric and can therefore be completely described by the effective elasticity tensor $\Cmacro$.

\subsection{Transverse 1D waves}

Assuming one-dimensional transverse wave propagation, the problem reduces to a single active deformation mode. Following the same procedure as before, we can write
\begin{align}
	\mumacro  =\murel \frac{\mumicro-\eta \, \omega^2}{\murel+\mumicro-\eta \, \omega^2} ,
\end{align}
with the two limits
\begin{align}
	\begin{aligned}
		\lim_{\omega \to 0} \mumacro = \frac{\murel \mumicro}{\murel + \mumicro}, \quad \quad \lim_{\omega \to \infty} \mumacro= \murel,
	\end{aligned}
\end{align}
This means that at very high frequencies, the microstructure cannot follow the fast oscillations, so the response is dominated by the effective shear stiffness $\murel$. 

The transition between these two limiting regimes is governed by the internal inertia of the microstructure. In particular, the effective modulus exhibits a zero at
\begin{align}
	\omega_z
	=
	\sqrt{\frac{\mu_{\mathrm{micro}}}{\eta}},
\end{align}
and a pole at
\begin{align}
	\omega_p
	=
	\sqrt{\frac{\mu_{\mathrm{rel}}+\mu_{\mathrm{micro}}}{\eta}}.
\end{align}
The vanishing of $\mumacro(\omega)$ at $\omega_z$ corresponds to a complete cancellation of the effective shear stiffness, whereas the pole at $\omega_p$ represents the natural resonance frequency of the internal microstructural degree of freedom. This means that, the effective modulus cannot be understood as an interpolation between $\mu_{\mathrm{macro}}$ and $\mu_{\mathrm{rel}}$, but exhibits a resonant transition controlled by the micro-inertia $\eta$. In practical materials this singular behavior is regularized by dissipation or viscosity, yielding a finite resonant peak instead of an ideal mathematical pole.

\subsection{Longitudinal 1D waves}

In 1D and for longitudinal waves, we get
\begin{align}
	\CCmacro 
	= \CCe \, 
	\frac{\CCmicro - \eta \, \omega^2  }
	{\CCe + \CCmicro - \eta \, \omega^2 } ,
	\label{eq:Cmacro_final_P_waves}
\end{align}
where for keeping generality, we have denoted $\CCmacro=2\mumacro+\lambdamacro, \CCe=2\mue+\lambdae, \CCmicro=2\mumicro+\lambdamicro$ denote the scalar projections of the corresponding fourth-order tensors onto the considered longitudinal mode.  The frequency-dependent longitudinal modulus exhibits a zero at
\begin{align}
	\omega_z
	=
	\sqrt{\frac{\CCmicro}{\eta}},
\end{align}
and a pole at
\begin{align}
	\omega_p
	=
	\sqrt{\frac{\CCe+\CCmicro}{\eta}}.
\end{align}
Note that both the effective longitudinal and transverse moduli possess the same frequency-dependent rational structure, characterized by a zero and a resonance pole. This behavior is a direct consequence of the dynamic condensation of the internal microdeformation field and is therefore independent of the particular deformation mode.

\subsection{Physical interpretation of elastic parameters}

Within the next sections, we aim to clarify the type of homogenization that it is aimed within the micromorphic framework and to draw distinctions with conventional static homogenization techniques. To this end, we analyze each elastic material parameter in detail, with the goal of providing a physically meaningful interpretation of the whole framework.

\subsubsection{The general static picture}

In the static case, the identification of material parameters for enriched continua is still today a highly non-trivial and largely unresolved problem with no universally accepted homogenization strategy accepted. Several methodologies have been previously proposed to upscale the heterogeneous model with microstructure to certain generalized model. In particular, homogenization towards Cosserat media has been addressed in \citep{hutter2019micro,jeong2009numerical}, and homogenization procedures involving the classical Eringen--Mindlin micromorphic continuum have been explored in \citep{alavi2021construction,biswas2017micromorphic,hutter2017homogenization,hutter2019micro,forest2002homogenization,rokovs2019micromorphic,rokovs2020extended,rokovs2020reduced,zhi2022direct}. However, all these studies raise fundamental questions regarding the identification and interpretation of additional material parameters, most of them often within the framework of first-order computational homogenization \citep{trinh2012evaluation,ganghoffer2023frontiers}.

Within the relaxed micromorphic framework, a fine identification procedure has been previously proposed. In \cite{neff2020identification}, it is shown that the tensor $\Crelax$ (in their study called $\mathbb{C}_{\rm micro}$) can be identified as the L\"owner half-order matrix supremum of suitable apparent stiffness tensors at the microscale. This means that $\mathbb{C}_{\rm micro}$ can be understood as the stiffest admissible elastic response of the unit cell under affine Dirichlet boundary conditions. On the other hand, the macroscopic tensor $\Cmacro$ follows from standard periodic homogenization and determines the mesoscopic elasticity tensor $\Crel$ through the exact micro--macro consistency condition (eq.~\eqref{eq.Consistency_Condition}). Both $\mathbb{C}_{\rm micro}$ and $\Cmacro$ are thus be obtained via first-order numerical homogenization at the unit-cell level \citep{neff2020identification,sarhil2023size,Agostino2022,sarhil2024computational,sky2024higher,sky2025novel}. This interpretation shows consistent limiting behavior. For a homogeneous unit cell, the identity $\mathbb{C}_{\rm micro} = \Cmacro$ implies $\Crel \to \infty$, and the micromorphic model reduces to classical linear elasticity. In contrast, the classical Eringen--Mindlin model degenerates into a second-gradient formulation with unbounded stiffness. We show here that this is false.

Within a different notation used for the relaxed micromorphic theory, note that the static tensorial consistency condition eq. \eqref{eq.Consistency_Condition}, can be written as follows
\begin{equation}
	\Cmacro = \left( \frac{1}{\Crel} + \frac{1}{\Crelax} \right)^{-1},
\end{equation}
where the macroscopic stiffness tensor $\Cmacro$ comes from the interaction between $\Crel$ and $\Crelax$, and formally resembles a harmonic-type combination at the scalar level \citep{neff2020identification,Sarhil2023,sarhil2023size,sarhil2024computational,sky2024higher,sky2025novel}.

\subsubsection{The general dynamical picture}
\label{sec.General_Dynamical_Regime}

In contrast to the previous presented static case, the dynamical regime introduces a frequency-dependent effective response due to micro-inertia effects. As a consequence, the macroscopic stiffness $\Cmacro$ cannot be described by a single constant elasticity tensor, but shows different frequency-dependent asymptotic limits. 

For both longitudinal and transverse polarizations, the macro--micro eq.~\eqref{eq.Cmacro_effective} leads to the two  limits
\begin{align}
		\lim_{\omega \to 0} \Cmacro = \Crel \, \Crelax \left(\Crel + \Crelax\right)^{-1}, \quad  \quad \quad \lim_{\omega \to \infty} \Cmacro= \Crel .
		\label{eq.micro_macro_limits}
\end{align}

The low-frequency limit $(\omega \to 0)$ recovers the static effective response (eq. \eqref{eq.Consistency_Condition}). In contrast, in the high-frequency regime, the macroscopic response is only governed by the elasticity tensor $\Crel$, that simply reflects the inability of the microstructure to fully relax. This transition between the two regimes highlights the role of internal length scales and micro-inertia in the micromorphic framework, which are absent in classical static homogenization approaches.

If we denote the lower limit $\displaystyle \lim_{\omega \to 0} \Cmacro = \Czero$ and the upper limit $\displaystyle \lim_{\omega \to \infty} \Cmacro = \Crel = \Cinfty$, we can write the following expression
\begin{align}
	\Crelax 
	=	\left(\Cinfty - \Czero\right)^{-1} \Czero \, \Cinfty-\eta \, \omega^2 \id .
	\label{eq.Crelax}
\end{align}

It is then simply to observe that $\Crelax$ represents a hidden scale between $\Czero$ and $\Cinfty$ with a pole at $\Czero = \Cinfty$. In other words, $\Crelax$ represents an intermediate scale governing the transition between the low-frequency and high-frequency responses, with a singular behavior as $\Czero \to \Cinfty$. 

Additionally, we can observe that eq. \eqref{eq.Crelax} encodes standard idea in rheology and viscoelasticity, i.e., the difference between $\Czero$ and  $\Cinfty$ originates from internal (micro) relaxation mechanisms. Within this interpretation,, $\Crelax$ measures the ability of the microstructure to relax and the term $\eta \,  \omega^2$ acts like a frequency-dependent stiffening. Consequently, $\Crelax$ controls the separation between the relaxed and unrelaxed responses, i.e., how far $\Czero$ lies below $\Cinfty$.

This formal analogy with classical relaxation-type constitutive models, in which the material response transitions between relaxed and unrelaxed moduli, can be illustrated with the standard linear solid (Zener) model of viscoelasticity. The frequency-dependent viscoelastic modulus can be written as follows \citep{Moczo2014}
\begin{equation}
	\Czero  = \Cinfty \, \frac{1 + i \, \omega \, \tau_\varepsilon}{1 + i \, \omega \, \tau_\sigma},
	\label{eq.Zener}
\end{equation}
which describes a transition between a relaxed modulus $\Czero $ and an unrelaxed modulus $\Cinfty$, governed by the strain relaxation time $\tau_\varepsilon$ and stress relaxation time $\tau_\sigma$ (related to the dissipative mechanisms of the material under study).

It is crucial to note that eq. \eqref{eq.Crelax} shows a similar mathematical structure to eq. \eqref{eq.Zener}, with $\Czero $ and $\Cinfty$ representing the low- and high-frequency limits, respectively. However, the underlying mechanism is fundamentally different in both cases; while the viscoelastic relaxation is associated with dissipative mechanisms and the loss of energy, the micromorphic frequency dependent behavior originates from micro-inertia rather than dissipation, resulting in a purely real and non-dissipative response.

\paragraph{Laboratory experiments}

\begin{figure}
	\begin{center}
		\includegraphics[width=1\textwidth]{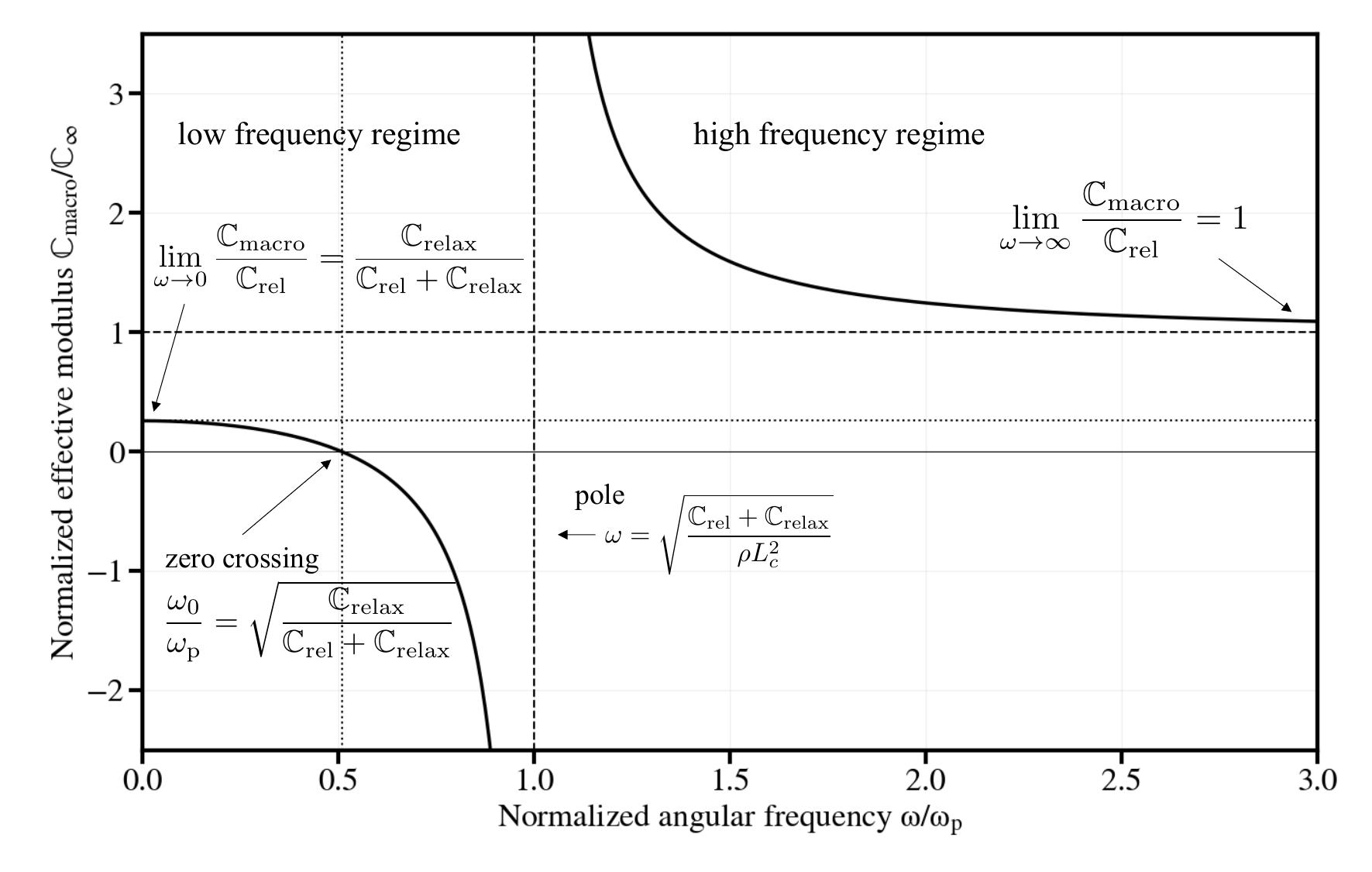}
		\caption{Normalized effective elastic modulus given by eq.~\eqref{eq.Cmacro_effective} as a function of the normalized angular frequency. The horizontal asymptotes correspond to the low- and high-frequency limits, while the vertical asymptote identifies the pole.}
		\label{Fig.micro_macro_scaling_variable}
	\end{center}
\end{figure}

The micro-macro eq.~\eqref{eq.Cmacro_effective}, reveals the two limiting frequency cases given by eqs. \eqref{eq.micro_macro_limits}. These limits give us information that can be obtained from laboratory experiments. At high frequencies we are able to directly obtain the elastic tensor $\Crel$ and a low frequencies the combination $\Crel \, \Crelax \left(\Crel + \Crelax\right)^{-1}$. From the low frequency expression, the value of $\Crelax$ can be readily determined after having obtained $\Crel$. This gives us a clear physical picture given in Fig. \ref{Fig.micro_macro_scaling_variable}, from which we can simply observe that the term $\Crelax$ can be understood as the internal stiffness associated with the microstructural relaxation responsible for the observed separation between the low- and high-frequency limits (eq. \eqref{eq.Crelax}).

\paragraph{The gap between high- and low-frequency moduli}

We define the frequency-dependent difference as
\begin{align}
	\Delta\mathbb C(\omega)
	:=
	\Cinfty-\Cmacro(\omega).
\end{align}
Since $\Cinfty=\Crel$, substitution of eq.~\eqref{eq.Cmacro_effective} and using $\eta=\rho \, L_c^2$ gives
\begin{align}
	\Delta\mathbb C(\omega)
	=
	\Crel
	\left(
	\Crel+\Crelax-\rho L_c^2\omega^2\id
	\right)^{-1}
	\Crel.
	\label{eq.Dimensional_Gap}
\end{align}
This frequency-dependent difference possesses a pole whenever
\begin{align}
	\Crel+\Crelax-\rho L_c^2\omega^2\id=0.
\end{align}
and it does not, in general, represent a smooth bounded transition between the low- and high-frequency elastic moduli.

The fixed gap between the two asymptotic limits is instead defined by
\begin{align}
	\Delta\mathbb C_{\rm asym}
	:=
	\Cinfty-\Czero
	=
	\Crel
	\left(
	\Crel+\Crelax
	\right)^{-1}
	\Crel.
\end{align}
Hence, $\Crelax$ controls the separation between the relaxed and unrelaxed asymptotic responses. In particular,
From which we can observe that 
\begin{itemize}
	\item \textbf{small $\Crelax \to 0$}: \, \, $\Delta\mathbb C_{\rm asym}\to \infty$ \(\Rightarrow\) large gap, strong relaxation.
	\item \textbf{large $\Crelax \to \infty$}: \, $\Delta\mathbb C_{\rm asym} \to 0$ \(\Rightarrow\) small gap, weak relaxation.
\end{itemize}

This shows that the asymptotic stiffness gap between the relaxed ($\Czero$) and unrelaxed ($\Cinfty$) elastic responses decreases as $\Crelax$ increases. Therefore, $\Crelax$ can be interpreted as the material parameter controlling the separation between the low- and high-frequency elastic moduli. Small values of $\Crelax$ correspond to a pronounced difference between the two asymptotic regimes (strong microstructural relaxation), whereas large values of $\Crelax$ reduce this difference until both limits eventually coincide. This interpretation is schematically illustrated in Fig.~\ref{Fig.micro_macro_scaling_variable}.

\paragraph{Interpretation of the relaxation parameter $\Crelax$}

Although the physical meaning of $\Crelax$ is clear, it can be obscure/unclear to understand what a large or small  $\Crelax$ will produce large or small gaps in practice. In simpler words, how do we know that we are in presence of a large or a small $\Crelax$, that can be physically translated into large or small relaxation? To answer this question and to gain complete physical insight, we next define, based on eq. \eqref{eq.Dimensional_Gap}, the dimensionless gap between high‑ and low‑frequency moduli as follows  
\begin{align}
 \left \langle \Delta \mathbb{C} , \Cinfty^{-1} \right \rangle=  \left \langle  
 \left(\Crel + \Crelax - \rho L_c^2 \omega^2 \id\right)^{-1}  , \Crel \right \rangle.
	\label{eq.Normalized_Gap}
\end{align}
This dimensionless gap $ \left \langle \Delta \mathbb{C} , \mathbb{C}_e^{-1} \right \rangle$ is a practical measure of the magnitude of relaxation at a given frequency relative to the instantaneous stiffness. By inspection of the normalized gap in eq.~\eqref{eq.Normalized_Gap}, one can define representative illustrative regimes of relaxation that provide a practical interpretation of the model response.  While no universal thresholds exist to classify the strength of relaxation, the normalized gap in Eq.~\eqref{eq.Normalized_Gap} allows representative regimes to be defined for a practical interpretation of the material response.

\subsubsection{The general dynamical picture: low frequency regime}

We next analyze the low frequency regime $(\omega \to 0)$, by neglecting the characteristic length $\mathbb{L}_c$ and $\Crot=\Cspin=0$ in the micromorphic energy (eq. \eqref{eq.General_Eringen_Micromorphic_Internal_Energy}), which reduces to 
\begin{align}
	 \mathcal{E}_{\rm total}  = \frac{1}{2} \underbrace{\left\langle \sy(\nabla u - P) ,\Crel \, \sy(\nabla u - P) \right\rangle}_{\text{relative deformation energy}} 
	 + \frac{1}{2} \underbrace{\left\langle \sy P , \Crelax \, \sy P \right\rangle}_{\text{relaxation resistance energy}} .
	\label{eq.Simplified_RM_Energy}
\end{align}

The tensor $\Crel$ governs the elastic energy associated with the mismatch between the macroscopic deformation $\nabla u$ and the micro-distortion $P$, through the relative deformation measure $\sy(\nabla u-P)$. In contrast, the tensor $\Crelax$ governs the energetic resistance to internal deformation modes by penalizing the activation of the micro-distortion field through the term $\sy P$.

\paragraph{Relationship between $\nabla u$ and $P$}

The minimization of the functional in eq. \eqref{eq.Simplified_RM_Energy} with respect to the field variables $(u,P)$, gives the following relation between $\nabla u$ and $P$
\begin{align}
P  = \Crel \left(\Crel+\Crelax\right)^{-1} \nabla u ,
	\label{eq.P_nablau}
\end{align}
which means that, in the static case, $P \le \nabla u$, i.e., the magnitude of $P$ is always proportional to the magnitude of $\nabla u$ and never larger. From eq. \eqref{eq.P_nablau}, we can find the next four limits
\begin{align}
\displaystyle \lim_{ \Crelax \to 0} P = \nabla u, \quad \quad \displaystyle \lim_{ \Crelax \to \infty} P = 0, \quad  \quad  \displaystyle \lim_{ \Crel \to 0} P = 0, \quad \quad  \displaystyle \lim_{ \Crel \to \infty} P = \nabla u  .
\end{align}
For fixed material parameters $(\Crel, \Crelax)$, three distinct regimes can be identified:
\begin{enumerate}
	\item \textbf{fully coupled regime:} $P\approx \nabla u$. The micro-distortion follows the macroscopic deformation, producing a negligible relative deformation, i.e., $\nabla u - P\approx 0$. This corresponds to the case where internal relaxation is almost absent $\Crelax \approx 0$.
	
	\item \textbf{suppressed microstructure:} $P \approx 0$. The microstructural deformation is inhibited, and the macroscopic deformation is almost entirely accommodated through the relative deformation. This corresponds to the case where internal deformation is strongly penalized by a large relaxation stiffness $\Crelax$.
	
	\item  \textbf{intermediate regime:} $P < \nabla u$. The internal deformation $P$ partially follows the macroscopic deformation $\nabla u$, which leads to a finite mismatch between the two fields.
\end{enumerate}

These limits demonstrate that $\Crel$ and $\Crelax$ jointly determine how deformation is sheared between the macroscopic field $\nabla u$ and the internal micro-distortion field $P$. A small $\Crelax$ allows the microstructure to closely follow the macroscopic deformation, whereas a large $\Crelax$ suppresses the development of internal deformation. Similarly, $\Crel$ controls the energetic cost associated with the mismatch between the two fields, thereby enforcing their kinematic compatibility. In contrast to the static case, the dynamic case emerges from the competition between the mismatch resistance governed by $\Crel$ and the relaxation resistance governed by $\Crelax$. This balance is further modified by micro-inertia, which determines whether the local deformation $P$ can respond to changes in the macroscopic deformation $\nabla u$ over a different frequency regime.

\subsubsection{The general dynamical picture: high frequency regime}

Including the kinetic energy of the local deformation in eq. \eqref{eq.Simplified_RM_Energy}, and setting $L_c=\Crot=\Cspin=0$, leads to the following expression
\begin{align}
	\mathcal{E}_{\rm total}  = \frac{1}{2} \underbrace{\left\langle \sy(\nabla u - P) ,\Crel \, \sy(\nabla u - P) \right\rangle}_{\text{relative deformation energy}} 
	+ \frac{1}{2} \underbrace{\left\langle \sy P , \Crelax \, \sy P \right\rangle}_{\text{relaxation resistance energy}} + \underbrace{\frac{1}{2} \left\langle \partial_t P,\eta \, \partial_t P\right\rangle}_{\text{microstructure kinetic energy}}.
	\label{eq.Simplified_RM_Energy_inertia}
\end{align}
The minimization of the functional in eq. \eqref{eq.Simplified_RM_Energy_inertia} with respect to the field variables $(u,P)$, and assuming plane wave solutions, gives the following frequency-dependent relation between $\nabla u$ and $P$ 
\begin{align}
\widehat{P} = \Crel \left(\Crel + \Crelax - \eta \, \omega^2 \id\right)^{-1} \, \nabla  \widehat{u},
\end{align}
where  $\widehat{P}$ and $\widehat{u}$ denote the amplitudes of the corresponding plane waves for $P,u$ respectively. In the high-frequency limit $\eta \, \omega^2 \id \gg \Crel + \Crelax$, we obtain
\begin{align}
	\widehat{P} \approx \nabla  \widehat{u} \;\;\to\; 0,
\end{align}
which means that the local deformation $P$ is strongly suppressed by inertia. Within this regime, the microstructure cannot follow the macroscopic deformation and its contribution thus becomes negligible. This behavior indicates that, at high frequencies, deformation is carried almost entirely by the macroscopic field $u$, leading to an effective response governed by the  stiffness $\Crel$.

\subsubsection{The characteristic length: $L_c\to \eta \, \partial_t^2 P$} 

The characteristic length $L_c$ defines an intrinsic length scale associated with the inertial response of the microstructure through the relation $\eta=\rho L_c^2$. Therefore, $L_c$ controls the dynamic resistance to acceleration and as a consequence, it does not represent a directly measurable geometric variable such as the grain size, but rather as a constitutive length scale governing the inertial response of the internal deformation modes.

It influence in wave propagation emerges through the combined scaling with frequency $L_c^2 \omega^2$, which represents a competition between microstructural inertia and elastic stiffness. In particular, one can observe dispersion effects when the wavelength becomes comparable to the effective length scale associated with $L_c$, or equivalently when $L_c^2 \omega^2$ becomes comparable to the total stiffness $(\Crel+\Crelax)$, see eq. \eqref{eq.Cmacro_effective}. When this happens, the propagating wave interacts dynamically with the microstructure, leading to frequency-dependent effective properties. 

Thus, $L_c$ should be interpreted as the constitutive length scale that determines the frequencies (or wavelengths) at which the inertial response of the microstructure becomes significant.

\subsubsection{The takeaway}

The micromorphic model can be interpreted as a form of frequency-dependent homogenization, in which the effective macroscopic stiffness emerges from the dynamic interaction between the macroscopic deformation $(\nabla u)$ and internal/local deformation $(P)$. The model does not attempt to identify the microstructure itself, but rather its mechanical effect on the macroscopic response. Unlike classical static homogenization, where effective properties depend solely on the geometry of the microstructure, the frequency dependence of the micromorphic model arises from the inertia and dynamics of the microstructural/local deformation field $P$. As a result, the material exhibits a dispersive effective response without requiring geometric constraints. In this sense, the model provides a dynamical average of the microstructural dynamics, where the effective properties depend on frequency due to the presence of internal inertia.

\begin{figure}
	\begin{center}
		\includegraphics[width=1\textwidth]{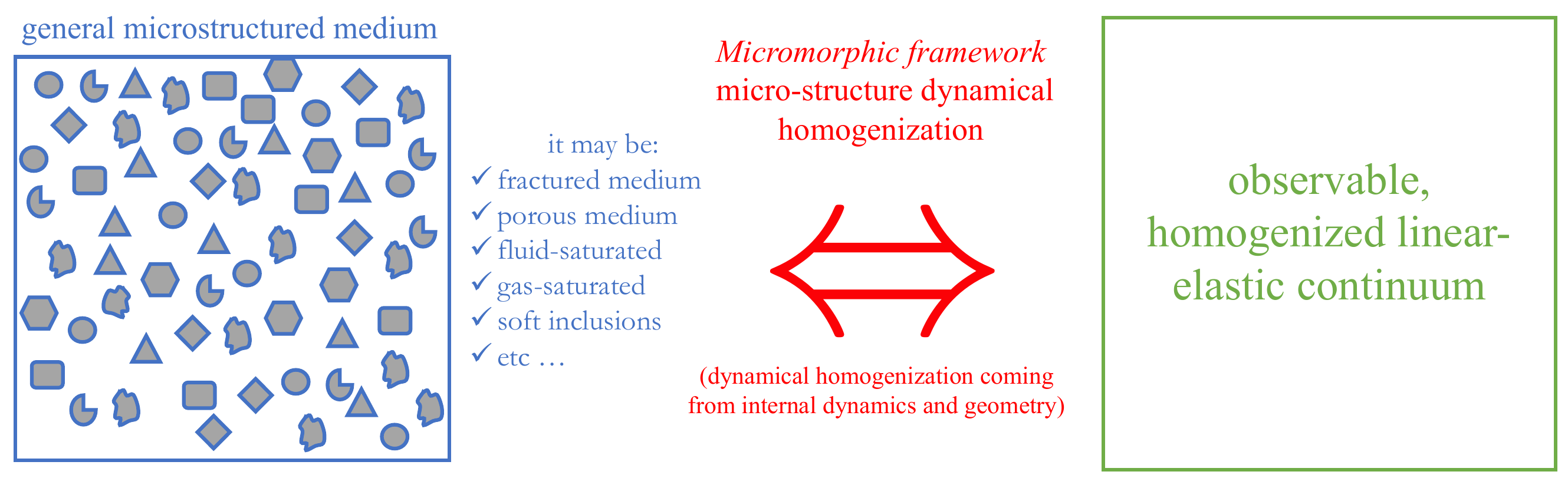}
		\caption{Different microstructural configurations (e.g., cracks, pores with fluid/gas filled regions, or hard/soft inclusions) can lead to indistinguishable macroscopic responses within the micromorphic framework. The model captures the effective stiffness of the internal deformation mechanism rather than the specific physical nature of the microstructure.}
		\label{Fig.relaxed_micro_homogenization}
	\end{center}
\end{figure}

The elastic tensors $\Crel$ and $\Crelax$ do not represent literal phase stiffnesses. Instead, they quantify the energetic resistance of macroscopic deformation versus microstructural/local deformation. The effective macroscopic stiffness $\Cmacro$ emerges from the dynamic coupling between $\nabla u$ and $P$, leading to frequency-dependent behavior and wave dispersion (see Fig. \ref{Fig.relaxed_micro_homogenization}).

An important consequence is that different microstructural configurations such as fractured media, porous media with fluid- or gas-saturated regions, soft/hard inclusions, etc., many may lead to the same effective response $\Cmacro$ (see Fig. \ref{Fig.relaxed_micro_homogenization}). Within this framework, $\Crelax$ should be interpreted as a relaxation parameter: it quantifies the resistance of the internal microstructural deformation and the amount of stress that can be relaxed or accommodated by the microstructure. Microstructural features that are compliant, soft, or capable of rearrangement all contribute to a lower $\Crelax$, while stiffer or more constrained internal structures produce a higher $\Crelax$. This means that $\Crelax$ does not directly identify the physical composition neither geometry of the microstructure but it provides an effective measure of the collective relaxation capacity of the microstructure.

\section{Non--linear shear softening}
\label{se_GeneralExperimentalValidations}

Several soils and/or rocks with different type of microstructure may undergo shear softening in presence of moderate to strong shear deformation. Shear softening simply refers to the reduce of the effective shear stiffness of the material. Capturing this behavior is of primordial importance for predicting phenomena such as liquefaction, strain localization, or damping in geomaterials \citep{hardin1965nature, lemaitre2992course, ishihara1996soil}. However, despite its importance and important consequences in seismological applications \citep[e.g.][]{hartzell2004prediction,Esfahani2024,ji2021effects,gueguen2019comparison,gelis20122,Lai2025}, the theoretical modeling of these kind of effects are commonly done relying on purely empirical hyperbolic or polynomial softening laws \citep{hardin1972shear,pyke1979nonlinear,hartzell2004prediction,beresnev1996nonlinear,ishihara1996soil,darendeli2001development}.  

Our goal in this section is to capture fundamental physics of how microstructural mechanisms such as grain rearrangements, contact degradation, and micro-crack activation progressively reduce the material shear stiffness under increasing shear strain. As we will see, unlike purely empirical hyperbolic laws, we aim to link macroscopic softening directly to fundamentally physical meaningful microstructural mechanics, allowing for a possible more realistic (or less empirical) representation of material behavior under large shear deformation. To do so, we next assume that the average shear properties of the material suffers any type of mechanical degradation as follows \citep{lemaitre2992course}
\begin{align}
	\mu(\gamma)=\mu ^0 \, (1 - d(\gamma)),
	\label{eq.mu_micro_d}
\end{align}
where $\mu ^0$ is the initial average shear modulus and we introduce a scalar variable $d(\gamma)$ as a function of the average shear strain $\gamma$ that represents the average shear degradation of the material as follows \citep{lemaitre2992course}
\begin{align}
	0 \le d(\gamma) \le 1,
\end{align}
where $d = 0$ is the intact material and $d = 1$ is fully degraded material. In physical terms $d$ may represent crack density, contact degradation, grain debonding, and/or micro-slip activation, etc \citep{lemaitre2992course}.

A physically reasonable evolution law is an exponential activation, consistent with classical continuum damage mechanics formulations \citep{lemaitre2992course}, as follows
\begin{align}
	d(\gamma) =	1 - \exp\!\left(-\frac{\gamma}{\gamma_c}\right),
\end{align}
where $\gamma_c$ is a characteristic shear strain controlling, for example, crack activation. Substituting $d(\gamma)$ into eq. \eqref{eq.mu_micro_d} we can write a strain dependent shear elastic parameter as follows
\begin{align}
	\mu (\gamma) = \mu ^0 \exp\!\left(-\frac{\gamma}{\gamma_c}\right).
	\label{eq.mu_micro_Softening}
\end{align}

We next interpret the shear modulus $\mu(\gamma)$ defined in eq. \eqref{eq.mu_micro_d} as the effective modulus measured at the scale of laboratory experiments. Such measurements are typically performed at relatively high frequencies, for which internal microstructural mechanisms do not have sufficient time to relax. As a result, in our interpretation, $\mu(\gamma)$ represents a high-frequency (unrelaxed) response of the material, in contrast to the lower-frequency behavior observed at seismic/seismological scales, where internal relaxation processes may significantly reduce much more the effective stiffness. 

Within the micromorphic framework, this high-frequency response is identified with the elastic coupling modulus $\murel$. Therefore, we set $\murel(\gamma) = \mu(\gamma)$, and the micromorphic consistency condition (eq.~\eqref{eq.General_Consistency_Condition}) can be written as follows
\begin{align}
	\mumacro(\gamma) = \frac{\murelax \, \left[\murel^0
		\exp\!\left(-\nicefrac{\gamma}{\gamma_c}\right) -  \eta \, \omega^2  \right]} {\murelax + \murel^0
		\exp\!\left(-\nicefrac{\gamma}{\gamma_c}\right) - \eta \, \omega^2 },
		\label{eq.Relaxed_MIcro_Soft1}
\end{align}
which produces nonlinear softening behavior. Note that for small strain and low frequencies $(\gamma,\omega \to 0)$, we obtain
\begin{align}
	\lim_{\gamma ,\omega \to 0} \mumacro= \frac{\murel^0 \, \murelax}
	{\murel^0 + \murelax}.
\end{align}
Thus, the small-strain stiffness is finite and physically meaningful. For shear large strains $(\gamma \to \infty)$ and low frequencies $(\omega\to 0)$ we have
\begin{align}
		\lim_{\gamma \to \infty,\omega\to 0} \mumacro= 0,
\end{align}
thus the material losses shear stiffness due to, for example, progressive crack growth. In physical terms, we can say that shear strain activates micro-cracks and crack growth reduces microstructure stiffness. The resulting reduced micro stiffness lowers the effective macroscopic modulus and as a consequence, softening emerges from internal structural degradation rather than from a fitted empirical formula, as we will next see.

Note that the strain-dependent internal resonance frequency in eq. \eqref{eq.Relaxed_MIcro_Soft1} is given by
\begin{align}
	\omega_p(\gamma)
	=
	\sqrt{
		\frac{
			\murelax
			+
			\murel^0
			\exp \left(-\frac{\gamma}{\gamma_c}\right)
		}{\eta}
	}.
	\label{eq.strain_dependent_resonance}
\end{align}
This frequency corresponds to the natural oscillation of the internal microstructural degree of freedom. Since the coupling modulus $\murel(\gamma)$ decreases with increasing shear strain, the resonance frequency also progressively shifts toward lower frequencies. As a consequence, progressive microstructural degradation not only reduces the effective shear stiffness but also lowers the characteristic frequency of the internal resonance. In the absence of dissipation, this resonance appears as a pole of the dynamically condensed effective modulus. 

\subsection{Relationship with the Hardin--Drnevich modulus reduction law}

The shear modulus decay with shear strain measured in the laboratory, is often represented with the following phenomenological Hardin--Drnevich modulus reduction curve equation \citep{hardin1972shear,pyke1979nonlinear,hartzell2004prediction,beresnev1996nonlinear,ishihara1996soil}
\begin{align}
	\frac{\mu}{\mu_0} = \frac{1}{1 + \gamma / \gamma_\text{ref}}
	\label{eq.Hyperbolic_Law}
\end{align}
where $\mu_0$ is the undisturbed shear modulus, $\gamma$ = macroscopic shear strain and $\gamma_\text{ref}=\tau_0/\mu_0$ is the reference strain controlling the onset of softening and $\tau_0$ is the maximum shear stress that the material can support in the initial state. 

To improve the fit to experimental data, several modifications to  eq. \eqref{eq.Hyperbolic_Law} have been proposed such as, Darendeli's single parameter $(\alpha)$ model and a two-parameter or modified Darendeli's model \citep{darendeli2001development,zhang2005normalized} given by the following expressions
\begin{align}
	\frac{\mu}{\mu_0} = \frac{1}{1 +\displaystyle \left( \frac{\gamma }{\gamma_\text{ref}}\right)^{\alpha}}, \quad \quad \quad \frac{\mu}{\mu_0} = \frac{1}{1 + \beta\displaystyle \left( \frac{\gamma }{\gamma_\text{ref}}\right)^{\alpha}} ,
\label{eq.Hyperbolic_Laws}
\end{align}
where $\alpha,\beta$ are phenomenological fitting parameters.

The set of phenomenological eqs. \eqref{eq.Hyperbolic_Law}--\eqref{eq.Hyperbolic_Laws} are simple and, in general, are able to match experimental observations. However, they are completely phenomenological and no physical interpretation of the microstructure is given. As an attempt to solve these limitations, we next introduce a micromorphic formulation framework based in eq. \eqref{eq.Relaxed_MIcro_Soft1}. To do so, we consider the regime of moderate strains and expand the exponential term in a first-order Taylor approximation in eq. \eqref{eq.mu_micro_Softening} as follows
\begin{align}
	\murel(\gamma) = \murel^0 \exp\!\left(-\frac{\gamma}{\gamma_c}\right) \approx \murel^0
	\frac{1}{1+\gamma/\gamma_c}.
\end{align}

Substituting this approximation into the  consistency condition eq. \eqref{eq.Relaxed_MIcro_Soft1}, leads to the following expression
\begin{align}
	\frac{\mumacro}{\murel^0}
	=
	\frac{\murelax}{\murel^0}
	\frac{
		\murel^0
		-
		\eta\omega^2
		\left(
		1+\displaystyle\frac{\gamma}{\gamma_c}
		\right)
	}{
		\murel^0
		+
		\left(
		\murelax-\eta\omega^2
		\right)
		\left(
		1+\displaystyle\frac{\gamma}{\gamma_c}
		\right)
	}.
	\label{eq.Mu_micro_Hyperbolic}
\end{align}

Note that eq. \eqref{eq.Mu_micro_Hyperbolic} is simply the ratio between the low frequency limit $\mumacro$ and high frequency limit $\mu_e$  (see Section \ref{sec.General_Dynamical_Regime}). Eq. \eqref{eq.Mu_micro_Hyperbolic} has the next two frequency limits 
\begin{align}
	\lim_{\omega \to 0}\frac{\mumacro }{\murel^0} =\frac{\murelax}{\murel^0+\murelax}
	\frac{1}{1+ \frac{\murelax}{\murel^0+\murelax}\gamma/\gamma_c}, \quad  \quad \lim_{\omega \to \infty} \frac{\mumacro }{\murel^0} = \frac{\murelax}{\murel^0 } .
\end{align}
At low frequencies, the term $\nicefrac{\murelax}{(\murel^0 + \murelax)}$ sets the \emph{maximum relative macroscopic modulus}, i.e., the value of $\mumacro/\murel^0$ at zero strain ($\gamma = 0$). Taking the limit of vanishing shear strain, $\gamma\rightarrow0$, in eq.~\eqref{eq.Mu_micro_Hyperbolic} gives
\begin{align}
	\lim_{\gamma\rightarrow0}
	\frac{\mu_{\rm macro}}{\mu_e^0}
	=
	\frac{\mu_{\rm relax}}
	{\mu_e^0+\mu_{\rm relax}-\eta\omega^2}
	\left(
	1-
	\frac{\eta\omega^2}{\mu_e^0}
	\right),
\end{align}
which is finite. In this limit, $\mu_e \rightarrow\mu_e^0$ (see eq.~\eqref{eq.mu_micro_Softening}), and the energetic penalty associated with the relative deformation $(\nabla u-P)$ reaches its maximum value. As a consequence, the micromorphic model tends to minimize the relative deformation between the macroscopic and microstructural kinematics. The extent to which this occurs, however, is controlled by the competition between the coupling stiffness $\mu_e^0$ and the relaxation stiffness $\mu_{\rm relax}$.

It is important to emphasize that $\mu_{\rm relax}$ does not represent the stiffness of a particular microstructural phase. Instead, it quantifies the energetic resistance for activating the internal microdeformation field. Large values of $\mu_{\rm relax}$ suppress the development of $P$, whereas smaller values allow the microstructure to accommodate a larger fraction of the imposed deformation.

Within this interpretation, the Hardin--Drnevich empirical law (eq.~\eqref{eq.Hyperbolic_Law}) may be regarded as the macroscopic manifestation of a progressive activation of internal relaxation mechanisms governed by the strain-dependent coupling parameter $\mu_e(\gamma)$.  Within the micromorphic framework, relaxation is not governed by a softening of a material phase, but by the progressive release of kinematic constraints between coupled deformation mechanisms.

\begin{figure}
	\begin{center}
		\includegraphics[width=1\textwidth]{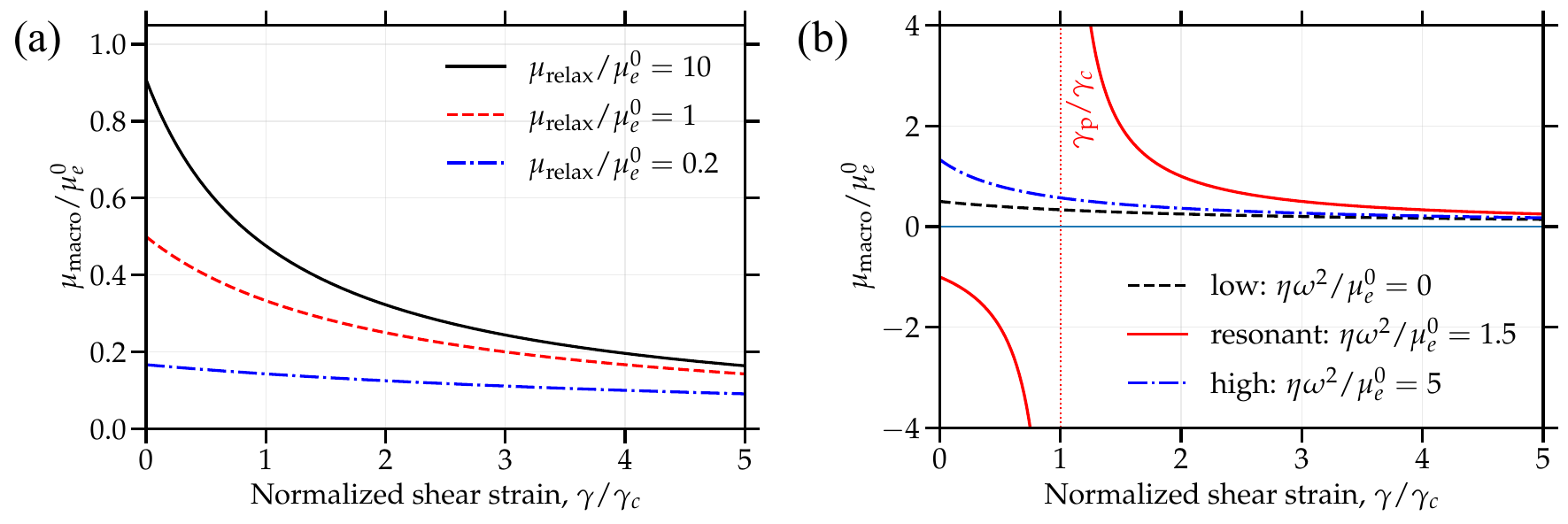}
		\caption{(a) Comparison of macroscopic softening curves. (b) Effect of frequency on the macroscopic softening response.}
		\label{fig.Relaxed_Softening}
	\end{center}
\end{figure}

In Fig.~\ref{fig.Relaxed_Softening}--a, when $\murelax \gg \murel^0$ (blue curve), the material starts close to its unrelaxed stiffness. In this regime, the large relaxation stiffness strongly penalizes the activation of the internal microdeformation field, so that $P \ll \nabla u$. As a consequence, only a limited amount of the
applied deformation is accommodated by the microstructure, and the macroscopic response remains comparatively stiff.

When $\murel^0 \sim \murelax$ (red dashed curve in Fig.~\ref{fig.Relaxed_Softening}--a), the internal microdeformation allows a larger fraction of the imposed
deformation to be relaxed internally. The effective macroscopic modulus is therefore reduced, and the strain-dependent softening becomes more pronounced.

Figure~\ref{fig.Relaxed_Softening}--b illustrates the influence of frequency. The micro-inertial contribution modifies the dynamic coupling
between the macroscopic deformation and the internal microstructure. As the excitation frequency increases, the effective modulus evolves
toward the characteristic internal resonance of the micromorphic medium. In the case where no viscoelastic effects are present, the effective modulus first decreases, vanishes at the characteristic frequency $\omega_z=\sqrt{\murelax/\eta}$, and subsequently exhibits a pole at $\omega_p=\sqrt{(\murel^0+\murelax)/\eta}$. Beyond this resonant regime, the microdeformation field is no longer able to follow the macroscopic motion, and the material response approaches the unrelaxed coupling stiffness $\murel(\gamma)$.

The resonant behavior predicted by the present formulation is a direct consequence of the dynamics of $P$, i.e., $\partial_t^2 P$. However, in real scenarios,  dissipative mechanisms are expected to regularize this singular behavior, producing a smooth transition between low and high frequency regimes..

Equation~\eqref{eq.Mu_micro_Hyperbolic} shows the interplay between the strain-dependent coupling stiffness, the resistance to internal relaxation, and the frequency-dependent inertial effects associated with the internal degrees of freedom. The general physical picture can be summarized as follows:
\begin{itemize}
	\item The term
	\begin{align}
		\frac{\murelax-\eta\omega^2}
		{\murel^0+\murelax-\eta\omega^2}
	\end{align}
	determines the normalized effective stiffness in the limit of vanishing shear strain. At low frequencies, it reflects the competition between the coupling stiffness $\murel^0$ and the resistance to internal deformation $\murelax$. The micro-inertial contribution modifies this balance as the frequency approaches the characteristic internal frequencies of the material.
	
	\item The hyperbolic-like term
	\begin{align}
		\frac{1}{
			1+
			\displaystyle
			\frac{\murelax-\eta\omega^2}
			{\murel^0+\murelax-\eta\omega^2}
			\frac{\gamma}{\gamma_c}
		}
	\end{align}
	governs the reduction of the effective modulus with increasing shear strain. This behavior originates from the strain-dependent degradation of the coupling stiffness $\murel(\gamma)$. The frequency modifies the apparent softening rate through the internal micro-inertia.
	
	\item In the low-frequency limit, $\omega\to0$, the normalized effective modulus becomes
	\begin{align}
		\lim_{\omega\to0}
		\frac{\mumacro}{\murel^0}
		=
		\frac{\murelax}
		{\murel^0+\murelax}
		\frac{1}{
			1+
			\displaystyle
			\frac{\murelax}
			{\murel^0+\murelax}
			\frac{\gamma}{\gamma_c}
		}.
	\end{align}
	In this regime, the internal degree of freedom can follow the macroscopic deformation and partially relax the imposed shear strain. The effective response is therefore softer than the unrelaxed coupling response.
	
	\item In the high-frequency limit, $\omega\to\infty$, the microdeformation amplitude tends to zero and
	\begin{align}
		\lim_{\omega\to\infty}
		\mumacro 
		=
		\murel(\gamma).
	\end{align}
	The internal deformation is dynamically constrained, and the material approaches the unrelaxed coupling stiffness. Unlike the previous interpretation, however, the strain dependence does not disappear in this limit because the coupling modulus itself has been assumed to degrade according to $\murel(\gamma)$.
	
	\item In general terms, increasing frequency suppresses the internal relaxation associated with the microdeformation field and shifts the response from the relaxed modulus toward the unrelaxed coupling modulus.
	
\end{itemize}

Equation~\eqref{eq.Mu_micro_Hyperbolic} therefore provides a description of the frequency- and strain-dependent material response. The model predicts a transition from a relaxation-dominated regime at low frequencies to an unrelaxed elastic regime at high frequencies. In the ideal lossless case, this transition is characterized by an internal resonance associated with the microstructural degree of freedom.

\subsection{Experimental verification}

We next match experimental observations to the introduced softening law (eq. \eqref{eq.Mu_micro_Hyperbolic}) using laboratory data from \cite{dammala2019dynamic}. The soil studied is a sand collected from the shore of Brahmaputra River near Guwahati city in Assam, India (a highly active seismic region). Based on the grain size distribution, the collected sand falls in the highly liquefiable materials category, classified as poorly graded sand according to unified soil classification system (USCS). The studied dynamic soil properties of these samples are  useful for performing seismic design of structures or seismic re-qualification of existing structures in similar soils \citep{dammala2019dynamic}.

Two types of apparatus (resonant column-RC and cyclic triaxial-CTX) were used to test the strain dependent dynamic properties and liquefaction potential of the samples. For our analysis, we use the variation of $\mu/\mu_0$ of the samples at 50$\%$ relative density and 50 kPa confining pressure. The reference strain value was considered from the experimental results at  $\mu/\mu_0$ of 0.5. 

To match the experimental data with the micromorphic model, we write eq. \eqref{eq.Mu_micro_Hyperbolic} as follows
\begin{align}
\frac{\mumacro}{\murel^0} =  
A \; \frac{ 1 - C \left( 1 + \frac{\gamma}{\gamma_c} \right) }{ 1 + B \frac{\gamma}{\gamma_c} },
\end{align}
where the effective parameters are defined as
\begin{align}
	A &= \frac{\murelax}{\murelax + \murel^0 - \eta \, \omega^2}, \quad \quad B= \frac{\murelax - \eta \, \omega^2}{\murelax + \murel^0 - \eta \, \omega^2} , \quad \quad C = \frac{\eta \, \omega^2}{\murel^0}.
\end{align}

\begin{figure}
	\begin{center}
		\includegraphics[width=1\textwidth]{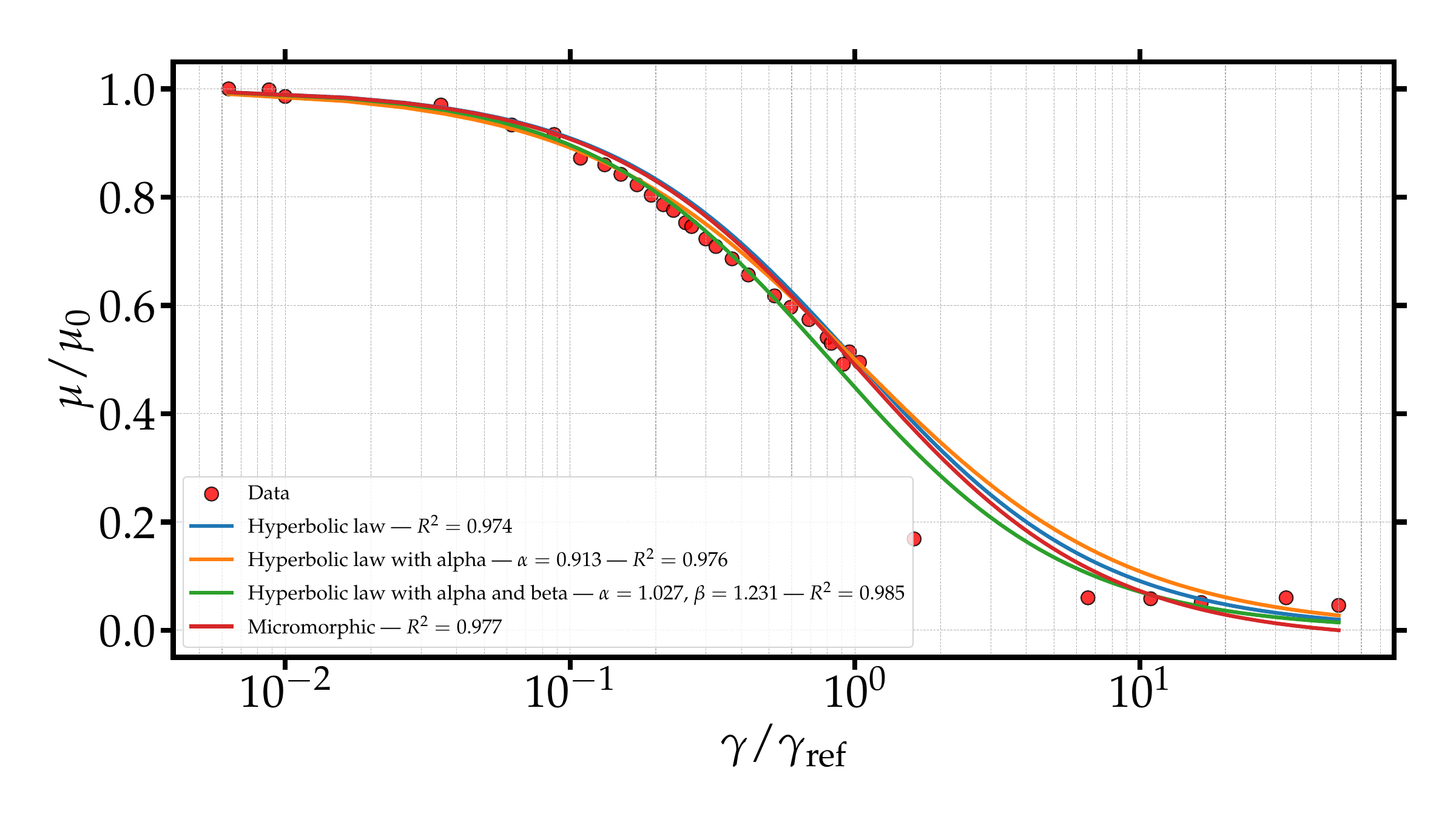}
		\caption{Soil data from the Brahmaputra River at 50$\%$ relative density and 50 kPa confining pressure \citep{dammala2019dynamic}, fitted against the micromorphic model (eq. \eqref{eq.Mu_micro_Hyperbolic}) and all phenomenological hyperbolic laws (eqs. \eqref{eq.Hyperbolic_Law}--\eqref{eq.Hyperbolic_Laws}).}
		\label{fig.Models_Softening}
	\end{center}
\end{figure}

The results of fitting the micromorphic (eq. \eqref{eq.Mu_micro_Hyperbolic}) and all hyperbolic (eqs. \eqref{eq.Hyperbolic_Law}--\eqref{eq.Hyperbolic_Laws}) models to the experimental data are shown in Fig. \ref{fig.Models_Softening}. The best micromorphic fitted parameters are 
\begin{align}
	A \simeq 1 , \quad \quad B \simeq 1, \quad \quad C = 0.0196.
	\label{eq.relaxed_micro_soft_paramaters}
\end{align}
Note that the values of $A$ and $B$ have been rounded for presentation. However, their not-rounded values satisfy the consistency relation $A=B+C(1-B)$.

Unlike traditional hyperbolic phenomenological laws, which provide purely empirical fits to experimental data (see Fig.~\ref{fig.Models_Softening}), the micromorphic softening law (eq.~\eqref{eq.Mu_micro_Hyperbolic}) allows for a physically grounded interpretation of the fitted parameters $(A,B,C)$:
\begin{itemize}
	\item $A \simeq 1$ indicates that the resistance to internal relaxation is large compared to both the coupling stiffness $\murel^0$ and the inertial contribution $\eta \, \omega^2$ $(\murelax
	\gg
	\murel^0-\eta\omega^2)$. This means that, in the present sub-resonant regime where $\eta\omega^2<\murelax$, internal deformation modes are weakly activated, and the material behaves close to its maximum stiffness at small strains.
	
	\item The fitted value $C=0.0196$ shows that $\nicefrac{\eta\omega^2}{\murel^0}=0.0196$. This means that the inertial contribution $\eta \, \omega^2$ is small under the tested conditions. This corresponds to a quasi-static (weakly dynamic) regime in which dynamic effects do not significantly influence the activation of deformation induced by microstructure, and far from the characteristic zero and pole predicted by the ideal lossless model. 
	
	\item $B \approx 1$ indicates that $\murelax-\eta\omega^2$ is large compared with $\murel^0$. Additionally, the difference between $A$ and $B$ is
	\begin{align}
		A-B
		=
		\frac{\eta\omega^2}
		{\murelax+\murel^0-\eta\omega^2}.
	\end{align}
	Because the micro-inertial contribution is small in the present experiment, $A$ and $B$ take very similar values. The parameter $B$ describes the combined influence of the resistance to internal deformation and micro-inertia on the strain-dependent denominator of the effective modulus.
\end{itemize}

The fitted parameters therefore indicate that the experimental response corresponds to a sub-resonant regime characterized by a strong resistance to internal microdeformation and a comparatively small micro-inertial contribution. The observed modulus reduction is mainly governed by the strain-dependent degradation of the coupling stiffness $\murel(\gamma)$, rather than by proximity to an internal resonance.

\section{Discussion}
\label{sec.Discussion}

\subsection{Softening laws}

The micromorphic hyperbolic law eq. \eqref{eq.Mu_micro_Hyperbolic} shows that nonlinear softening originates from the progressive reduction of the coupling shear stiffness $\murel(\gamma)$, which governs the interaction between macroscopic deformation and internal degrees of freedom. In this sense, $\murel(\gamma)$ determines how strongly the macroscopic deformation constrains the local deformation. The elastic tensor $\murelax$ represents a relaxation parameter controlling the resistance to internal deformation modes. Material softening therefore arises from a progressive weakening of this constraint, allowing internal deformation mechanisms to become increasingly active. In addition, the characteristic length $L_c$ governs nonlocal and dynamic effects. The term $2 \, \eta \, \omega^2$ suppresses internal relaxation at high frequency, linking nonlinear softening with dispersion within a single model.

Within this physical picture, nonlinear softening is not empirically imposed, but naturally emerges from the interplay between coupling mechanisms and relaxation. At very small strains, $\murel \approx \murel^0$ and internal modes are strongly constrained, leading to a stiff response. As shear strain increases, $\murel(\gamma)$ decreases, progressively enabling internal relaxation mechanisms associated with phenomena such as grain rearrangement, contact slip or crack opening to contribute more
strongly to the effective macroscopic response.

Different microscopic mechanisms may lead to different evolutions of the coupling stiffness $\mu_e(\gamma)$. As a consequence, the present micromorphic formulation does not prescribe a unique softening mechanism, but provides a common mechanical framework within which different microstructural processes may be represented through their influence on the coupling between macroscopic and internal deformation. In other words, the model captures the effective response of internal mechanisms without requiring explicit specification of the underlying microstructure.

Although the inertial contribution is negligible under the experimental conditions investigated here, the formulation naturally extends to dynamic loading, where micro-inertia becomes significant and the model predicts the emergence of an internal resonance. In real materials this resonant behavior is expected to be regularized by dissipative mechanisms. The effective fitting parameters $(A,B,C)$ are not purely empirical quantities, but can be directly related to the coupling stiffness, the relaxation stiffness and the micro-inertial contribution. In this sense, the fitting process provides a direct mechanical interpretation of the observed modulus reduction curves.

From this perspective, classical modulus reduction curves do not constitute independent constitutive laws, but rather represent low-frequency macroscopic manifestations of a more general micromorphic relaxation process.

\subsection{Dynamic homogenization}

Note that in general, the expressions derived for elastic parameters reveal a mathematical structure that can be written in the canonical form as follows
\begin{align}
	Q_{\rm macro}
	= Q_{\rm micro}\,
	\frac{\text{internal dynamics}}
	{\text{external coupling} + \text{internal dynamics}},
\end{align}
where $Q$ denotes a generic constitutive quantity. This structure reflects the underlying physics of the micromorphic framework. The macroscopic response emerges from a competition between internal microstructural dynamics and the stiffness of the surrounding medium. In particular, the frequency dependence enters through the term $L_c^2 \omega^2$, which governs the activation of internal deformation mechanisms. 

Such a structure naturally arises in dynamic homogenization theories. In contrast to classical static homogenization, where the internal degrees of freedom are eliminated and only their average effect is retained, the micromorphic framework incorporates them explicitly, leading to frequency-dependent effective properties. A simple illustration of this mechanism is provided by mass-in-mass systems, for which the effective dynamic mass takes the form \citep{huang2009negative}
\begin{align}
	M_{\rm eff}(\omega)=M + m\,\frac{\omega_0^2}{\omega_0^2 - \omega^2},
	\label{eq.general_resonator}
\end{align}
where $M$ is the macroscopic mass and $m$ represents an internal oscillator with natural frequency $\omega_0$. The rational dependence on frequency arises from the interaction between macroscopic motion and internal dynamics \citep{Srivastava2015_review}.

Similarly, in elastodynamic homogenization theories, such as the framework introduced by Willis \citep{willis1980polarization,willis1980polarizationb,willis1981variational,willis1985nonlocal}, the effective constitutive relations become frequency dependent and include coupling terms between momentum and strain fields \citep{nassar2015willis}. More recent developments show that both frequency and wavenumber dependence naturally emerge when microscale dynamics are retained in the upscaled description \citep{Meng2018_Willis,BLESGEN2026105561}. The micromorphic model provides a particularly transparent and physically interpretable dynamic homogenization. Rather than deriving effective parameters through asymptotic expansions, the model introduces internal degrees of freedom directly at the continuum level. Thus, the local field $P$ should be understood as an effective continuum representation of the collective dynamics of the microstructure rather than as the motion of a specific microscopic constituent. As a result, dispersion and frequency-dependent behavior emerge naturally from the governing equations, without requiring explicit assumptions about the underlying microstructure.

\subsection{Implications for distributed acoustic sensing (DAS) measurements}

Distributed Acoustic Sensing (DAS) measures axial strain averaged over a finite gauge length $L_g$, as follows \citep[e.g.][]{kennett2022seismic}
\begin{align}
	\varepsilon^{\text{DAS}}(x)
	=
	\frac{1}{L_g}
	\int_{x-L_g/2}^{x+L_g/2}
	\partial_s u(s)\, ds .
\end{align}
In practice, the averaged strain rate is obtained by differentiating the ground velocity resolved along the cable at the ends of the gauge length $L_g$ \citep{bakku2015fracture}. In numerical wave propagation highly complex and heterogeneous media, the gradient $\nabla u$ may exhibit strong spatial fluctuations or even do not mathematically exists $\nabla u \notin C^0(\Omega)$. To solve this problem, classical homogenization approaches are applied to regularize the gradient calculation  \citep[e.g.][]{capdeville2024sensitivity} and/or the Biot's theory of poroelasticity has been explored in realistic seismological scenarios of forward and inverse problems \citep[e.g.][]{masson2006finite,li2011wave,dupuy2016downscaling,couples2019phenomenological,yang2021frequency,liu2023time,wang2024petrophysical}.

The presented micromorphic framework provides a dynamic counterpart of classical homogenization techniques used to model sub-wavelength heterogeneity effects \citep[e.g.][]{capdeville2024sensitivity} and to the highly ill-posed inverse problem to determine Biot's elastic parameters \citep[e.g.][]{wang2024petrophysical}. While on the one hand classical homogenization replaces micro-structure by an effective static medium, the micromorphic model retains additional internal degrees of freedom, which provide a natural regularization mechanism to the effective strain $(\nabla u - P)$, and on the other hand, the model naturally accounts for both spatial averaging and frequency-dependent effects, allowing to properly reproduce the fundamental physics of wave propagation in poroelastic media. This offers a complete different physically grounded avenue to explore DAS measurements within highly heterogeneous media. 

This is also relevant for near-surface geophysics, where the seismic record is also mainly dominated by surface waves. From a theoretical and numerical perspective, surface waves represent a complex subject because their interaction with highly heterogeneous media leads to very complex waveforms. As a consequence, their accurate modeling is crucial to properly image the interior of the Earth \citep{fichtner2010full}. Recent efforts have been taken in this direction in order to put surface waves in micro-continuum media into a rigorous mathematical foundation \citep{KHAN2022102898,BULGARIU2024112661,Apetrii2025}.

\section{Conclusions}
\label{sec.Conclusions}

The micromorphic framework provides a unified and physically consistent description of materials with micro-structure. It is capable of simultaneously capturing nonlinear softening, frequency-dependent wave propagation, and scale effects within a single model. Unlike classical phenomenological approaches, the observed macroscopic behavior does not need to be prescribed a priori, but emerges naturally from the interaction between macroscopic deformation $\nabla u$ and internal degrees of freedom $P$.

We have shown that nonlinear softening can be understood as a consequence of the progressive reduction of the coupling between macroscopic deformation and internal local deformation. At small strains, the coupling stiffness is large and internal local deformation is constrained, leading to a stiff response. As strain increases, this constraint weakens, allowing internal relaxation mechanisms to become progressively active and reducing the effective macroscopic stiffness. In this sense, classical modulus reduction curves can be understood as macroscopic manifestations of an underlying relaxation process governed by internal kinematics.

From a theoretical perspective, the micromorphic framework can be interpreted as a kinematically enriched effective medium that extends classical static homogenization techniques. Rather than eliminating the microscale, it retains internal degrees of freedom that encode unresolved deformation mechanisms. Within this framework, the parameters do not correspond to distinct material phases, but to effective contributions controlling coupling, relaxation, and inertia. 

Overall, the proposed framework provides a parsimonious yet powerful description of complex material behavior. With a small number of parameters, it captures nonlinear softening, dispersion, and scale-dependent effects without requiring explicit knowledge of the underlying microstructure. This makes the framework particularly well suited for realistic applications, where the detailed geometry of the medium is often unknown. More generally, the model offers a new perspective in which macroscopic material behavior is governed by the activation and suppression of internal relaxation mechanisms, rather than by the properties of distinct phases, providing a unified basis for interpreting experimental observations across scales.

From this perspective, classical modulus reduction laws are not viewed as independent constitutive relations, but as particular macroscopic limits of a more general micromorphic relaxation theory. This provides a direct mechanical interpretation of empirical softening laws while naturally extending them to include nonlocality, dispersion, and frequency-dependent effects within a unified constitutive framework.

\section{Acknowledgments}

R.A. acknowledges comments of Patrizio Neff on earlier versions of this paper and constructive conversations with Fabian Bonilla. 

\section{Data availability}

The experimental data used for model validation were digitized from the published results of \cite{dammala2019dynamic}. No new experimental data were generated in this study. All equations required to reproduce the theoretical results are provided in the manuscript.

\section{Conflict of interest}

The authors have no conflicts of interest.

\footnotesize

\bibliographystyle{apalike}
\bibliography{Biblio}

\end{document}